\documentclass[a4paper]{jpconf}
\usepackage{graphicx}
\begin{document}
\title{Non-Fermi liquid transport and ``universal" ratios in  quantum Griffiths phases}

\author{David Nozadze and  Thomas Vojta}

\address{Department of Physics, Missouri University of Science $\&$ Technology, Rolla, MO 65409, USA}

\ead{dn9z2@mst.edu, vojtat@mst.edu.}

\begin{abstract}
We use the semi-classical Boltzmann equation to investigate transport properties such as
electrical resistivity,  thermal resistivity, thermopower, and the Peltier coefficient
of  disordered metals close to  an antiferromagnetic quantum phase transition. In the
quantum  Griffiths phase, the electrons are scattered  by spin-fluctuations in the rare regions. This
leads to singular temperature dependencies not just at the quantum critical point,
but in the entire Griffiths phase. We show that the resulting  non-universal
power-laws in transport properties are controlled by the same Griffiths
exponent $\lambda$ which governs the thermodynamics. $\lambda$ takes the value
zero at the quantum critical point and increases throughout the Griffiths phase.
We also study some of the ``universal" ratios commonly used to characterize Fermi-liquid
behavior.
\end{abstract}

\section{Introduction}

In recent years, quantum phase transitions \cite{Sachdevbook}  have  attracted a lot of attention
in condensed matter theory.
Quantum phase transitions happen at absolute zero temperature. They are induced  by the change of an external
parameter (pressure, chemical composition, and magnetic field) and are driven by quantum fluctuations.
A number of metallic systems show strong deviations from   conventional Landau Fermi-liquid properties when
they are tuned through a quantum critical point \cite{1994_Lohneysen_PRL}.

Significant  attention has also been attracted by phase transitions in the presence of
quenched disorder. It has become clear that at low temperatures, strongly correlated
materials can show a surprising sensitivity  to imperfections and disorder.
Interplay between large-scale quantum fluctuations and random fluctuations due to disorder
leads to exotic phenomena such as quantum Griffiths singularities \cite{1969_Griffiths_PRL, 1995_Thill_Physyca_A, 1996_Rieger}, infinite randomness
critical points featuring exponential instead of power-law scaling \cite{1992_Fisher_PRL, 1995_Fisher_PRB} and the smearing of
the phase transition \cite{2003_Vojta_PRL}. The Griffiths effects are caused by
large spatial
regions (rare regions) that are devoid of impurities and can show local order even if the bulk system is in
the disordered phase. The locally ordered rare regions are not static but retain
their quantum dynamics. Griffiths showed that these rare regions can  lead to a singularity in the free energy in a whole
parameter region which is now known as the Griffiths phase. In a \emph{quantum} Griffiths phase, the rare-region low-energy density of states
follows a power law, $\rho(\epsilon)\propto\epsilon^{\lambda-1}$, where $\lambda$
is the non-universal Griffiths exponent. $\lambda$ varies systematically  within the
Griffiths phase and vanishes at the critical point. This kind of density of states leads to power-law dependencies
of several observables on the temperature \textit{T}, including
 specific heat, $C\sim T^\lambda,$ and  magnetic
susceptibility, $\chi\sim T^{\lambda-1}$.  The zero-temperature magnetization-field curve
behaves as $M\sim H^{\lambda}$ (for reviews, see Refs.\ \cite{2006_Vojta_JPhysA, 2010_Vojta_JLTPhys}).
While the thermodynamics of quantum Griffiths phases is comparatively well understood,
much less is known about transport properties.

We showed in Ref.\ \cite{2011_Nozadze_EPL} that the rare-region contributions to electrical resistivity, thermal resistivity,
thermopower, and the Peltier coefficient in the quantum Griffiths phase associated
with an antiferromagnetic quantum phase transition are characterized by non-universal
power-laws in $T$ which are controlled  by the same Griffiths exponent $\lambda$
which also governs the thermodynamics.
Here, we summarize  these results.  We then investigate the behaviors
in the antiferromagnetic quantum Griffiths phase of some of
the ``universal"  ratios commonly used to characterize Fermi-Liquid behavior in metals.
Our paper is organized as follows.
In Sec. 2 we briefly discuss the model and  methods of solutions.
In Sec. 3  we then summarize the   derivations of the transport properties and
find that the scattering of the  electrons  by spin-fluctuations in the rare regions leads to singular temperature dependencies
not just at the quantum critical point but
in the entire antiferromagnetic quantum Griffiths phase. In Sec. 4 we present the behaviors of various ``universal" ratios.
Finally, we conclude in Sec. 5.

\section{Model}

The transport properties of the itinerant  antiferromagnetic systems we are interested in can be described by
a two-band model consisting of $s$ and $d$ electrons \cite{1966_Mills_JPhys, 1975_Ueda_JPhysSJ}. The Hamiltonian has the form
\begin{equation}
H=H_{s}+H_{d}+H_{s-d}\,,
\end{equation}
 where $H_{s}$ and  $H_{d}$ are the Hamiltonians of  $s$ and $d$ electrons, respectively.
Only the $s$ electrons contribute to the transport properties. They are scattered
by the spin-fluctuations of the $d$ electrons which are assumed to be in the antiferromagnetic quantum
Griffiths phase. The contribution to the  resistivity  from the scattering by the spin-fluctuations stems from
the $s-d$ exchange interaction term of the Hamiltonian
\begin{eqnarray}
H_{s-d}=g \int d  \textbf{r} \ \textbf{s}(\textbf{r})\cdot \textbf{S}(\textbf{r})\,,
\end{eqnarray}
where $g$ is the  coupling between $s$ and $d$ electrons. \textbf{s} and
\textbf{S} are the  spin densities of the $s$ and $d$ electrons, respectively.

Near the quantum critical point in three dimensions,
the concept of quasiparticles is still  (marginally)
well defined, therefore transport properties can be treated
within a semi-classical Boltzmann approach.
The linearized Boltzmann equation in the presence of a temperature gradient $\nabla T,$ and
 an electric field $\textbf{E}$  but zero magnetic field can be written as \cite{Zimanbook}
\begin{eqnarray}
-\textbf{v}_{\bf{k}}\frac{\partial f^0_{\bf{k}}}{\partial T} \nabla T
-\textbf{v}_{\bf{k}}\frac{\partial f^0_{\bf{k}}}{\partial \varepsilon_{\bf{k}}}
 \textbf{E} =\frac{\partial f_{\bf{k}}}{\partial t}\bigg{\vert}_{\rm{scatt}} \label{BE}\,,
\end{eqnarray}
where $f^0_{\bf{k}}$  is the equilibrium Fermi-Dirac distribution function.
The first and second  terms describe  changes  of the electron distribution
function $f_{\bf{k}}$ due to  diffusion and electric field $\textbf{E}$, respectively.
The last one is the collision  term.
Let the stationary solution of the Boltzmann equation be $f_{\bf{k}}=f^0_{\bf{k}}-\Phi_{\bf{k}}
(\partial f^0_{\bf{k}}/\partial \varepsilon_{\bf{k}}),$
where $\Phi_{\bf{k}}$ is a measure of the deviation  of the electron distribution from equilibrium. Then the linearized
scattering term due to the spin-fluctuations  has the form \cite{1975_Ueda_JPhysSJ, 1995_Hlubina_PRB}
\begin{eqnarray}
\frac{\partial f_{\bf{k}}}{\partial t}\bigg{\vert}_{\rm{scatt}}&=&\frac{2g^2}{T}\sum_{\bf{k}'}f^0_{{\bf{k}}'}(1-f^0_{\bf{k}})n(\varepsilon_{\bf{k}}-\varepsilon_{\bf{k}'})
 \rm{Im} \chi({\bf{k}}-{\bf{k}}', \varepsilon_{\bf{k}}-\varepsilon_{{\bf{k}}'})(\Phi_{\bf{k}}-\Phi_{{\bf{k}}'})\,
 \nonumber \\ &=& \frac{1}{T}\sum_{{\bf{k}}'} \mathcal{P}_{{\bf{k}}'}(\varepsilon_{\bf{k}}-\varepsilon_{{\bf{k}}'})(\Phi_{\bf{k}}-\Phi_{{\bf{k}}'})\label{Scatt}\,,
\end{eqnarray}
where $n(\varepsilon_{\bf{k}}-\varepsilon_{{\bf{k}}'})$ is the Bose-Einstein distribution function and
$\chi$ is the total dynamical susceptibility of the spin-fluctuations ($d$ electrons).
 In the above equation, the spin-fluctuations are
assumed to be in equilibrium. This approximation is valid if the system can lose momentum efficiently by
Umklapp or impurity scattering.

Quantum Griffiths effects in disordered metallic systems  are realized in both Heisenberg magnets \cite{2005_Vojta_PRB} and
 Ising magnets. In the latter case, they occur in a transient temperature range where the damping is unimportant \cite{2000_Castro_PRB}.
In the following, we consider both cases.

In order to study  the rare-region contribution to the transport properties in the Griffiths phase, we need to
find the total rare-region dynamical susceptibility. To get it,
we can simply  sum over the susceptibilities of the individual clusters (rare regions).
The imaginary part of the  dynamical  susceptibility of a single cluster of characteristic energy $\epsilon$
of a disordered itinerant quantum Heisenberg  antiferromagnet in the  quantum Griffiths phase  is given by
\begin{eqnarray}
\textrm{Im} \chi_{\rm{cl}}    (\mathbf{q},\omega; \epsilon )=\frac{\mu^2 \gamma \omega}{\epsilon^2(T)
+\gamma^2\omega^2}F_{\epsilon}^2(\mathbf{q})\label{ChiHeis}\,,
\end{eqnarray}
where $\mu$  is the moment of the cluster and $\gamma$ is the damping coefficient which results from the coupling of the spin-fluctuations
and the electrons.
 $\epsilon(T)$ plays the role of the local  distance from criticality. For high temperatures $\gamma T \gg \epsilon,$ $\epsilon(T)\approx T$ and for low temperatures
$\gamma T \ll \epsilon,$ $\epsilon(T)\approx \epsilon.$ $\textit{F}_{\epsilon}(\mathbf{q})$ is the form factor of the cluster
which encodes the spatial magnetization profile.

The imaginary part of the  dynamical  susceptibility of a single cluster in random quantum Ising models is given by
 \begin{eqnarray}
\textrm{Im} \chi_{\rm{cl}}(\mathbf{q},\omega; \epsilon )&=&\pi \frac{\mu^2}{4}\tanh \left (\frac{\epsilon} {2T} \right ) [\delta (\epsilon-\omega)- \delta (\epsilon+\omega)]F_{\epsilon}^2(\mathbf{q})\label{ChiIsing}\,.
\end{eqnarray}
The precise functional form of $F_{\epsilon}(\mathbf{q})$
is not known, but we can  find it approximately by
analyzing the Fourier transform of a typical  local magnetization profile of the rare region.
We find $F^2_{\epsilon}(\textbf{q})=X[(\mathbf{q}-\mathbf{Q})^3\log (\epsilon^{-1})] \label{F2}$,
where $X$ and $\mathbf{Q}$ are a scaling function and the ordering wave vector, respectively \cite{2011_Nozadze_EPL}.

 To estimate the total rare-region susceptibility, we integrate over all rare regions using the  density
of states $\rho(\epsilon),$
\begin{eqnarray}
\rm{Im} \chi(\mathbf{q},\omega )= \int_{0}^{\Lambda}d\epsilon \rho(\epsilon)  \rm{Im} \chi_{cl}(\mathbf{q},\omega; \epsilon)
\label{Chi1}\,,
\end{eqnarray}
where $\Lambda$ is an energy cut-off.
In the Heisenberg case we find  that the rare-region contribution to the zero-temperature susceptibility in the quantum Griffiths phase
 can be expressed as (up to logarithmic corrections)
\begin{equation}
\rm{Im} \chi(\mathbf{q},\omega) \propto |\omega|^{\lambda-1} \rm{sgn}(\omega)\, X[(\mathbf{q}-\mathbf{Q})^3\log (\omega^{-1})]\label{Chi2} \,.\label{Chi3}
\end{equation}
The rare-region susceptibility of the random quantum Ising model has the same structure as Eq.(\ref{Chi2}) \cite{2000_Castro_PRB}.

\section{Transport properties}

\subsection{Electrical resistivity}

In order  to calculate the electrical  resistivity we  consider Ziman's variational principle \cite{Zimanbook}.
The resistivity $\rho$  is given as the minimum of a functional of $\Phi_\textbf{k}$  \cite{Zimanbook},
\footnote{We set Plank's constant, electron's  charge and Boltzmann constant $\hbar=e=k_{B}=1$ in what follows.}
\begin{eqnarray}
\rho[\Phi_{\bf{k}}]=\rm{min}
\Biggl[\frac{1}{2T}\frac{\int\int(\Phi_{\bf{k}}-\Phi_{{\bf{k}}'})^2\Gamma^{{\bf{k}}'}_{{\bf{k}}}d{\bf{k}} d{\bf{k}}'}{\bigl(\int v_{\bf{k}} \Phi_{\bf{k}}\frac{\partial f^0_{\bf{k}}}{\partial \varepsilon_{\bf{k}}} d{\bf{k}}\bigr)^2}\Biggr]\,, \label{rhofunc}
\end{eqnarray}
where
\begin{eqnarray}
\Gamma^{{\bf{k}}'}_{{\bf{k}}}=
\int_{0}^{\infty}d  \omega\  \mathcal{P}_{{\bf{k}}'} (\omega)\delta(\varepsilon_{\bf{k}'}-\varepsilon_{\mathbf{k}}+\omega) \,,
\end{eqnarray}
with $\mathcal{P}_{{\bf{k}}'} (\omega)$ defined in Eq.(4).

By making a suitable ansatz for the distribution function $\Phi_{\bf{k}}$ in the
functional (\ref{rhofunc}), we can find the resistivity. Close to the
antiferromagnetic quantum phase transition, the susceptibility is strongly peaked
around the ordering wave vector $\textbf{Q}$. This leads to anisotropic
scattering processes. However, in the presence of a high concentration
of impurities,  the low temperature resistivity is dominated by elastic
impurity scattering which is isotropic. The isotropic scattering
redistributes  the electrons over the Fermi surface. This allows
us to make the standard ansatz
\begin{eqnarray}
\Phi_{\mathbf{k}} \propto \textbf{n}\cdot\textbf{k}\,,\label{Phi1}
\end{eqnarray}
where $\textbf{n}$ is a unit vector parallel to the electric field. Note that any constant prefactor in $\Phi_{\bf{k}}$ is unimportant because it drops out in the resistivity functional (\ref{rhofunc}).
Then, following the calculations  for the electrical resistivity outlined in Ref. \cite{Zimanbook}, we obtain
\begin{eqnarray}
\Delta \rho \propto T \int d^3\textbf{q} \frac{(\textbf{n}\cdot \textbf{q})^2}{q} \int_{0}^{\infty}
d\omega \frac{\partial n(\omega)}{\partial T} \rm{Im} \chi(\textbf{q},\omega)\,.
\end{eqnarray}

Inserting the susceptibility (\ref{Chi3}), the temperature dependence of the electrical resistivity due to the spin-fluctuations in the Griffiths phase is given by
\begin{eqnarray}
\Delta \rho  \propto  T^\lambda \,,
\end{eqnarray}
up to
logarithmic corrections.
Thus, the temperature-dependence of the electrical resistivity follows a non-universal power-law governed by the Griffiths exponent $\lambda$.

\subsection{Other transport properties}

In the same way, we study the thermal resistivity.
We find that
the rare-region contribution to the thermal resistivity
in the antiferromagnetic quantum Griffiths phase has the form
\begin{eqnarray}
\Delta W\propto T^{\lambda-1} \,.
\end{eqnarray}

The Seebeck effect, the existence of an electric field $\mathbf{E}$ in a metal subject to a thermal gradient $\nabla T$,
is characterized  by the thermopower $S$ which is defined by the relation $\mathbf{E}=S \ \nabla T $.
To calculate the thermopower, we analyze the Boltzmann equation (3) in the presence of both $\nabla T$
and $\mathbf{E}$. Elastic impurity scattering leads to the usual linear temperature dependence $S_{\rm{imp}}\propto T$ while the contribution due to the magnetic scattering
by the rare regions in the Griffiths phase reads
\begin{eqnarray}
\Delta S\propto   T^{\lambda+1} \,.
\end{eqnarray}

Another transport coefficient called the Peltier coefficient $\Pi$ characterizes the flow of
a thermal current in a metal in the absence of a thermal gradient. It is related to the thermopower
by $\Pi=S\ T.$ Correspondingly, the rare-region contribution to the Peltier coefficient has the form
\begin{eqnarray}
\Delta \Pi \propto \ T^{\lambda+2} \,.
\end{eqnarray}

\section{``Universal" ratios in  quantum Griffiths phases}

In this section, we investigate the behaviors in quantum Griffiths phases  of some of
the ``universal"  ratios commonly used to characterize Fermi-liquid behavior. Two of the ratios, the Wilson
ratio $\chi T/C$ \cite{1975_Wilson_RMP, 1984_Stewart_RMP} and the  Gr{\"{u}}neisen parameter \cite{1912_Gruneisen_AP},
the ratio between  thermal expansion coefficient $\beta$ and specific heat $C$,  only involve thermodynamics. They have been discussed before.
In the Wilson ratio, the Griffiths-phase power-laws of $C$ and $\chi$ cancel, leading to a logarithmic temperature dependence \cite{2005_Miranda_RPP}
\begin{equation}
R_{W}=\chi T/C \propto [\log(1/T)]^2\,.
\end{equation}

The   Gr{\"{u}}neisen parameter in the quantum Griffiths phase  also
diverges logarithmically with vanishing temperature \textit{T} \cite{2009_Vojta_PRB2} 
\begin{eqnarray}
\Gamma=\beta/C \propto\log(1/T)\,.
\end{eqnarray}

The knowledge of the transport properties (Sec. 3) allows us to study further ratios. According to
the Wiedemann-Franz law \cite{1853_Franz_AP}, the ratio of the electrical resistivity divided by the temperature and
 the thermal resistivity, the so-called Lorenz number
$L=\rho/WT$, takes the universal value $L_0$  within Fermi-liquid theory.
In the Griffiths phase, dominant impurity scattering leads to the  universal value $L_0$ while  the
rare-region contributions to the Wiedemann-Franz law  give a subleading contribution 
\begin{eqnarray}
\Delta L \propto T^{\lambda}\,.
\end{eqnarray}

Let us now consider the quasi-universal ratio $q=S/C$ of the Seebeck coefficient
to the specific heat \cite{2004_Behnia_JPhysCond}. Due to the spin-fluctuations in the Griffiths phase,
this ratio strongly deviates  from its usual temperature-independent behavior  and is given by
\begin{eqnarray}
q \propto T^{1-\lambda} \,.
\end{eqnarray}

Another ``universal" ratio is the  Kadowaki-Woods ratio, normally defined as $AT^2/C^2$ \cite{1986_Kadowaki_SSC},
where $A$ is the coefficient of the $T^2$-term of the resistivity. To evaluate it in the Griffiths phase
where the leading temperature dependence of the resistivity is not quadratic,
we rewrite the definition as $(\rho-\rho_{\rm{imp}})/C^2$.
Then, we find that the Kadowaki-Woods ratio behaves as
\begin{equation}
R_{KW}=(\rho-\rho_{\rm{imp}})/C^2\propto T^{-\lambda}\,
\end{equation}
in the antiferromagnetic quantum Griffiths phase, in contrast to the temperature-independent Fermi-liquid behavior.

\section{Conclusions}

We  have studied  the transport properties in  antiferromagnetic quantum
Griffiths phases. The rare-region contributions to electrical resistivity, thermal resistivity,
thermopower, and the Peltier coefficient are characterized by non-universal
power-laws in $T$ which are controlled  by the Griffiths exponent $\lambda$.
We have also investigated the behaviors of some ``universal" ratios. In  antiferromagnetic quantum
Griffiths phases, the Kadowaki-Woods ratio and
the ratio between Seebeck coefficient and specific heat show strong deviations from the usual
 Fermi-liquid behavior. In contrast, the Wiedemann-Franz law is fulfilled  to leading order
in $T$, but the spin-fluctuations in the rare regions  lead to a subleading   non-universal
power-law in $T$ controlled  by the Griffiths exponent.
The two ratios involving thermodynamics only, the Wilson ratio and the Gr{\"{u}}neisen ratio,
show only logarithmic deviations from Fermi-liquid behavior.

Our results have been obtained using the semi-classical Boltzmann equation approach. As discussed in Ref.\ \cite{2011_Nozadze_EPL}, this approach is valid in the  Griffiths phase and may break down sufficiently close to the
quantum critical point. We also emphasize that our results have been derived for
antiferromagnetic quantum Griffiths phases and may not be valid for ferromagnetic systems. The problem is that
a complete theory of the ferromagnetic quantum Griffiths phase in a metal does not exist. In particular, the
dynamical susceptibility is still  not known. This work is in progress.

\ack
This work has been supported by the NSF under Grant No. DMR-0906566.

\section*{References}

\providecommand{\newblock}{}

\end{document}